\documentclass[twocolumn,showpacs,amsmath,amssymb,nofootinbib]{revtex4}

\usepackage{graphicx}
\usepackage{dcolumn}
\usepackage{bm}

\usepackage{color}
\newcommand{\modif}[1]{#1}

\begin{document}


\title{Deposition of Na Clusters on MgO(001)}
\author{M. B\"{a}r$^1$, P.~M. Dinh$^{2,3}$, L.~V.~Moskaleva$^4$,
P.-G. Reinhard$^1$, N. R\"{o}sch$^4$, and E. Suraud$^{2,3}$}
\affiliation{
1) Institut f\"ur Theoretische Physik, Universit\"at Erlangen,
Staudtstrasse 7, 91058 Erlangen, Germany\\
2) Universit\'e de Toulouse; UPS; Laboratoire de Physique
  Th\'eorique (IRSAMC); F-31062 Toulouse, France\\
3) CNRS; LPT (IRSAMC); F-31062 Toulouse, France\\
4) Technische Universit\"at M\"unchen, 
Department Chemie and Catalysis Research Center, 
85747 Garching, Germany
}

\date{\today}

\begin{abstract}
We investigate the dynamics of deposition of small Na clusters on
MgO(001) surface. A hierarchical modeling is used combining
Quantum Mechanical with Molecular Mechanical (QM/MM) description.
Full time-dependent density-functional theory is used for the cluster
electrons while the substrate atoms are treated at a classical level.
We consider Na$_6$ and Na$_8$ at various impact energies.  We analyze
the dependence on cluster geometry, trends with impact energy, and
energy balance. We compare the results with deposit on the much
softer Ar(001) surface.
\end{abstract}

                             
\maketitle

\section{Introduction}

A major branch of present-days cluster research comprises clusters in
contact with solid surfaces, for an overview see, e.g.,
\cite{Mei00,Bin01,Mei06,MeiEPJD07}. \modif{The interaction of these
two entities gives rise to a rich scenery of effects such as, e.g.,
chemical reactions at surfaces 
\cite{surfacebasics2,surfacebasics3,surfacebasics5},
particularly, catalytic
applications \cite{surfacebasics1,surfacebasics4,San99},
or modified optical response\cite{Die02,Pin04a,Feh07a}.}
One crucial aspect here  is the process of
cluster deposition which is relevant for synthesis and for analysis of
clusters in contact with surfaces. Moreover, the deposition dynamics
as such is an interesting and demanding process due to the subtle
interplay of the impact of interface energy, electronic band structure of the
substrate, and surface corrugation.  
\modif{Accordingly, there is a wealth of investigations on cluster
deposition, experimentally oriented
\cite{Exp1,Exp2,Exp3,Bro96,Fed98a,Lau03a,Sie06,Duf07a}, theoretically
with molecular dynamics (MD) techniques
\cite{Xir02,CL,Hab93,Web07a,Jae07a} or more detailed quantum
mechanical methods
\cite{MH,Hak96b,Mos02a,Ipa03,Ipa04,cointrimers,Din07a}, for reviews
see \cite{Jen99aR,Bin01,surfacebasics1}.
\modif{Although addressing the same physical processes, 
these various theoretical
approaches, relying on different approximations, often provide useful
complementary information. }}
%
Recently,
we have investigated deposition dynamics of Na clusters on Ar(001)
surface \cite{Din07a,Din08a,Din09}. The aim of this paper is to
continue these 
theoretical studies now considering  deposition dynamics of Na clusters 
on a much "harder" surface than Ar, namely  MgO(001) insulator surfaces.
The structural properties and optical response of Na$_n$ on MgO have
already been studied in great detail in using the present
computational approach \cite{Bae07a}.  Both Ar and MgO are similar in
that they are both insulators with a large band gap, \modif{but they
differ significantly in other important properties.} There are,
however, large differences in other properties.  Ar is a Van-der-Waals
bound material, thus very soft with little surface corrugation. On the
other hand, MgO is an ionic crystal, well bound and with large surface
corrugation. It is thus most interesting to see how deposition
dynamics proceeds in that case, as such, and at variance with the Ar
case.

The theoretical description of clusters on surfaces is very involved
due to the huge number of degrees-of-freedom of these
systems.  \modif{This holds the more so for dynamics.  The vast
majority of theoretical studies thus resorts to MD simulations using
effective force fields between the atoms, as mentioned above
\cite{Xir02,CL,Hab93,Web07a,Jae07a,Jen99aR}.  These are comparatively
inexpensive and can provide a pertinent picture of the leading atomic
transport processes. Metal clusters are more than an ensemble of atoms
because they \modif{held together by delocalized bonds (due to a
common electron cloud); in consequence they show} pronounced shell
effects \cite{Hee93,Bra93,Bjo99}. This makes a quantum mechanical
description of cluster dynamics advisable.  Most of the fully quantum
mechanical pictures make compromises in concluding on dynamical
features from a series of static calculations.  A true
Born-Oppenheimer MD for deposition of Pd clusters on MgO substrate can
be found in \cite{Mos02a}. The enormous expense of such high
level calculations limits the size of the systems, particularly the
size of the representative for the substrate. On the other hand, there
are many situations in which the substrate is much more inert than the
cluster. Our test case of Na clusters on MgO(001) belongs to that
class. This suggests to use}
%
a hierarchical description where the cluster electrons are treated
quantum-mechanically by full Time-Dependent Density-Functional theory
(TDDFT) while the substrate atoms are handled at a lower level of
refinement by classical motion.  This modeling belongs to the family
of coupled Quantum-Mechanical with Molecular-Mechanical methods
(QM/MM) which are often used in other fields as, e.g., bio-chemistry
\cite{Fie90a,Gao96a,Gre96a} or surface physics
\cite{Mit93a,Nas01a}. In earlier studies, we developed and applied a
QM/MM model for Na clusters in contact with Ar
\cite{Ger04b,Feh06a,Feh05c,Din08a,Din09}. We have shown that it was
most crucial to 
include properly the dynamical polarizability of the substrate when
exploring truly dynamical processes as we aim at. Recently, we
extended the modeling to Na clusters on MgO surfaces, again including
dynamical polarizability \cite{Bae07a}. Here we take up that
model and apply it to a study of deposition dynamics. We will consider
Na$_6$ and Na$_8$ as test cases. These two clusters have very
different geometries and binding properties which allows to explore
qualitatively the impact of cluster properties on the deposition
process.

The paper is organized as follows:
In section \ref{sec:model}, we summarize the QM/MM model for
Na  clusters on  MgO.
Section \ref{sec:results} presents results tracking the detailed
dynamics in terms of trajectories and analyzing the processes with
respect to energy transfer and energy balance.
Conclusions are summarized in section \ref{sec:concl}.

\section{Brief summary of the model}
\label{sec:model}

\subsection{The degrees-of-freedom}

\begin{table}[ptb]%
\begin{tabular}
[c]{ll}\hline
$\varphi_{n}(\vec{r})\,,\,n=1,...N_{\mathrm{el}}$ & valence electrons of the
Na cluster\\
$\vec{R}_{i^{\mathrm{(Na)}}}\,,\,i^{\mathrm{(Na)}}=1...N_{\mathrm{i}}$ &
positions of the Na$^{+}$ ions\\\hline
$\vec{R}_{i^{(c)}}\,,\,i^{(c)}=1...M$ & positions of the O cores\\
$\vec{R}_{i^{(v)}}\,,\,i^{(v)}=1...M$ & center of the O valence cloud\\
$\vec{R}_{i^{(k)}}\,,\,i^{(k)}=1...M$ & positions of the Mg$^{2+}$
cations
\\
\hline
\end{tabular}
\caption{The dynamical degrees-of-freedom of the model. 
Upper block~: Na cluster. Lower
block~: Active cell of the MgO substrate. See text for details.}%
\label{tab:freedom}
\end{table}

The hierarchical QM/MM model has been detailed in \cite{Bae07a}. We
review that here briefly.  The various constituents and their
degrees-of-freedom are summarized in table \ref{tab:freedom}.
The Na cluster is treated in standard fashion \cite{Cal00,Rei03a}.
Valence electrons are described in terms of single-particle
wavefunctions $\varphi_{n}(\vec{r})$ and the complementing Na$^{+}$
ions are handled as charged classical point particles characterized by
their positions $\mathbf{R}_{i^{\mathrm{(Na)}}}$, see upper block of
table \ref{tab:freedom}.
\modif{The electrons are described by TDDFT
at the level of the local density approximation (LDA)}
The substrate is composed of two species: Mg$^{2+}$ cations and
O$^{2-}$ anions. The cations are electrically inert and can be treated
as charged point particles; they are labeled by $i^{(k)}$. The anions
are easily polarizable, an aspect which is described by allowing for two
constituents: a valence electron distribution (labeled by $i^{(v)}$)
and the complementing core (labeled by $i^{(c)}$). Each of these three
types of constituents is described as a classical degree-of-freedom in
terms of positions $\vec{R}_{i^{\mathrm{(type)}}}$, see the lower
block of table \ref{tab:freedom}. The difference $\mathbf{R}^{(c)}
-\mathbf{R}^{(v)}$ represents the electrical dipole moment of the
O$^{2-}$ anion and is thus allowed, by construction, to explicitly evolve 
in time, as a function of the local electric field due to 
all constituents of the system at a given instant. 

\begin{figure}[htbp]
\centerline{
\includegraphics[width=0.8\linewidth]{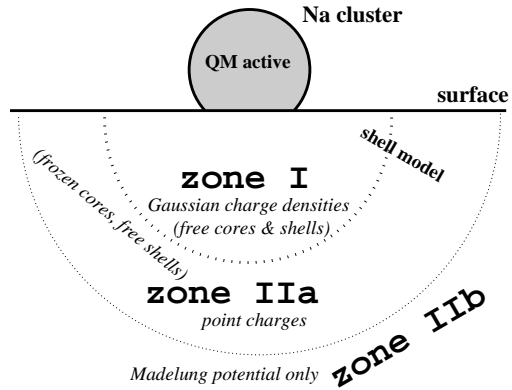}
}
\caption{\label{fig:model} Schematic view of the hierarchical
model for Na$_N$ on MgO(001) surface.  }
\end{figure}

The combined system is sorted in four stages of decreasing activity,
as sketched in figure ~\ref{fig:model}.  The Na cluster is
treated \modif{at the highest level of theory with} full TDLDA-MD.
The Mg and O ions of the substrate are arranged in \modif{fcc}
crystalline order corresponding to bulk MgO, \modif{with a lattice
parameter of 7.94 $a_0$}. All dynamical degrees-of-freedom for Mg
and O, as listed in table \ref{tab:freedom}, are taken into account in an
active cell of the MgO(001) surface region underneath the Na cluster,
denoted ``zone I'' in the sketch. The active cell is continued by an
outer region of MgO material (``zone IIa'') where the ionic centers of
Mg and O are kept fixed, while oxygen dipoles still remain active
degrees-of-freedom.  \modif{Thus zone I together with zone IIa
constitute the ``active cell''.}  Anything farther out (``zone IIb'')
is totally frozen at crystalline configuration and only its Madelung
potential is considered.  The effect of the outer region on the active
part is given by a time-independent shell-model potential
\cite{Dic58}; the actual parameters of this force field were adopted
from \cite{Nas01a}.
\modif{The active cell consists of three layers, each
containing square 
arrangements of 242 Mg$^{2+}$ cations and 242 O$^{2-}$ anions.  }
The ions in the lowest layer are fixed \modif{at the bulk
structure} to prevent them from relaxing and forming an artificial
second surface. The \modif{volume 
where ions and electrons are mobile (zone I)}
has a diameter of 24~$a_0$, all
\modif{layers} together \modif{(zone I+IIa)}
extend to 42~$a_0$ from the surface. \modif{Bulk structure farther
  out is modeled by the Madelung potential.}
\modif{Checks with models of a larger number of layers showed that
three layers provide an adequate description of the very
inert MgO material. (The soft Ar(001) substrate, used for comparison
later on, is more critical and requires at least four active
layers plus two frozen ones.)}

\subsection{The energy}

The total energy is composed as
$
E=E_{\mathrm{Na}}+E_{\mathrm{MgO}}+E_{\mathrm{coupl}}
$
where $E_{\mathrm{Na}}$ describes an isolated Na cluster,
$E_{\mathrm{MgO}}$ the MgO(001) substrate, and $E_{\mathrm{coupl}}$
the coupling between the two subsystems. For $E_{\mathrm{Na}}$, we
take the standard TDLDA-MD functional as in previous studies of free
clusters \cite{Rei03a,Cal00} including an average self-interaction
correction \cite{Leg02}. The energy of the substrate and the coupling
to the Na cluster consist of long-range Coulomb energy and some
short-range repulsion which is modeled through effective local
core-potentials \cite{Nas01a}.  To avoid the Coulomb singularity and
to simulate the finite extension of Mg$^{2+}$ and O$^{2-}$ ions, we
associate a smooth charge distribution
$\rho(\vec{r})\propto\exp{(-{\vec{r}^{2}}/{\sigma^{2}})}$ with each of
these ionic centers. We associate a similar smooth charge distribution 
to the O$^{2-}$ valence cloud as well. 
This altogether yields a soft Coulomb potential to be used
for all active particles.  

\modif{
\subsection{Calibration of the QM/MM model}

The calibration of the whole model has to address three issues:
The cluster as such, the environment as such, and the coupling between
both.
The modeling for the cluster is taken over from work on free
clusters \cite{Cal00,Rei03a}.
The model parameters for the pure environment are the same as
in previous studies of MgO(001) \cite{Nas01a,Win06}.
The parameters for the coupling between environment and Na cluster
were calibrated from scratch. The tuning for Na@MgO(001) was
performed using fully quantum-mechanically computed Born-Oppenheimer
surfaces for Na atoms and Na$^+$ ions on MgO(001) from
\cite{Win06}. These surfaces were computed at four different substrate
sites (O$^{2-}$, Mg$^{2+}$, hollow, bridge) down to close distances
where the full substrate repulsion was felt.
For further details and the actual model parameters, see \cite{Bae07a}.

Two quantitative points are worth to be mentioned. The modeling
achieves a barrier for penetration of cluster electrons into the
substrate which reproduces nicely the large band gap of 6.9 eV for
MgO. The fully quantum mechanical calculations show that electron
transfer from  the substrates O$^{2-}$ anions to a Na atom remains
below 0.1 charge units down to the closest distances considered
(where the repulsive energy comes about the band gap). Transfer from
the Na atom to the Mg$^{2+}$ cation is totally ignorable. This nice
decoupling of ad-atoms and substrate is probably a feature of simple
metals. Noble metals, e.g., can develop a more involved surface
chemistry due to the closeness of the $d$ shell \cite{Mos02a}.

}

\subsection{Solution scheme}

From the energy functional, once established, one derives the static
and dynamical equations variationally in a standard manner. 
The numerical solution of the coupled quantum-classical system
proceeds as described in \cite{Feh05a,Feh05b,Bae07a}. The
electronic wavefunctions and spatial fields are represented on a
Cartesian grid in three-dimensional coordinate space.  The numerical
box employed here has a size of $(64\, a_0)^3$.  The spatial
derivatives are evaluated via fast Fourier transform. The ground state
configurations were found by interlaced accelerated gradient
iterations for the electronic wavefunctions~\cite{Blu92} and simulated
annealing for the ions in the cluster and the substrate. Propagation
is done by the time-splitting method for the electronic
wavefunctions~\cite{Fei82} and by the velocity Verlet algorithm for
the classical coordinates of Na$^{+}$ ions and MgO constituents.

All the collisional processes studied in the current paper
proceed on an ionic time scale, i.e. slow as compared to electronic
motion.  True electronic excitations are thus extremely small.  For
example, ionization stays safely below a fraction of 0.001 electrons.
It would then be well justified to use Born-Oppenheimer-MD rather than
full TDLDA-MD, as long as one carefully maintains the crucial dipole
polarizability of the substrate.  But the TDLDA-MD scheme is so
efficient that it is still preferable for reasons of computing
time. Remind that the dipole response of the substrate needs to be
propagated at electronic time scale and dipole stepping is more
economic than fully relaxing the dipoles in each Born-Oppenheimer
step.

\subsection{Preparation of the system}

First, the ground state structures of the pure MgO surface and of the
free Na cluster are determined for the given model by simulated
annealing.  The Na cluster is then placed at a certain distance from
the surface of the substrate.  A distance of about $15\,a_0$ has
turned out to be sufficient. After that a Galilean transformation is
applied to the cluster.  This means that each of
the cluster ions is given a momentum $\vec P_0$ in the direction
towards the surface and the electronic wave functions are boosted by
an equivalent momentum as
\begin{equation}
\varphi_i(\vec r) \rightarrow \exp (\imath \vec p_0 \cdot\vec r)\,
\varphi_i(\vec r) 
\end{equation}
where $\vec p_0 = \frac{m_e}{M} \vec P_0$, $m_e$ and $M$ are the
electron and Na ion mass, respectively. This provides the initial
state from which on the system propagates in a straightforward manner
according to the TDLDA-MD equations.

\subsection{Structure of the test cases}
\label{sec:struct}

\begin{figure}[htbp]
\centerline{
\includegraphics[width=0.2\linewidth,angle=-92]{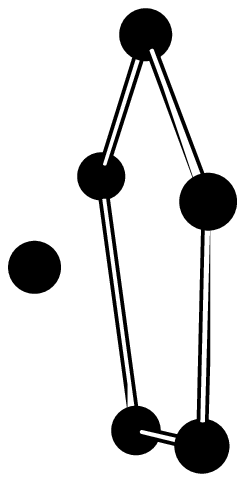}
\hspace*{2em}
\includegraphics[width=0.3\linewidth,angle=-90]{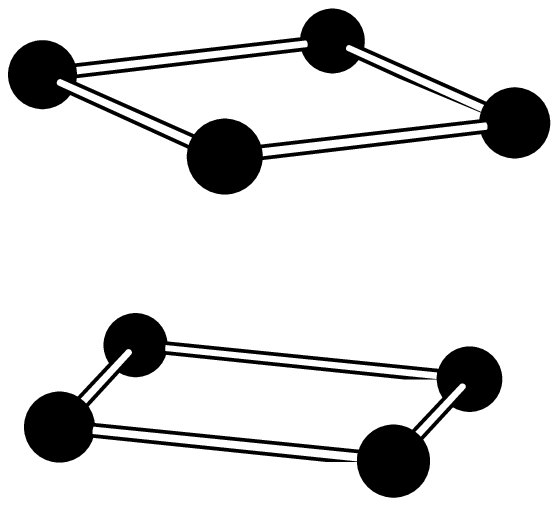}
}
\caption{\label{fig:struct}
The structures of free Na$_6$ (left) and Na$_8$ (right).
\modif{The bond distance along the five-fold ring of Na$_6$ is 6.5
  $a_0$ and the top ion resides 3.1 $a_0$ above the ring.
The bond distance in the two four-fold rings of Na$_8$
is 6.2 $a_0$ and the distance between the  two rings is 5.8 $a_0$.}
}
\end{figure}
The starting point of deposition dynamics are well relaxed
structures for the clusters and pure MgO(001) surface. These had been
discussed extensively in \cite{Bae07a}. The MgO surface is a cut
through cubic crystal structure. From the top, one sees a chess-board
structure with alternating Mg and O ions. For the Na clusters, we will
use here Na$_6$ and Na$_8$ as examples. The initial state starts from
free clusters. Their structures are shown in figure
\ref{fig:struct}. Note that the vertical axis in the figure will
represent the direction perpendicular to the surface in the
forthcoming deposition processes ($z$ axis).  Na$_6$ is strongly
oblate consisting out of a ring of five ions topped by one single
ion. Na$_8$ has a highly symmetric configuration out of two rings of
each four ions tilted relative to each other by 45$^\circ$ to minimize
Coulomb energy. The electronic cloud of Na$_8$ is close to spherical
shape because $N=8$ electrons correspond to a strong shell closure for
Na clusters~\cite{Bjo99}. It is important to note that the bond
distances for Na$_8$ \modif{(6.2 $a_0$)} are not far from the
\modif{diagonal} distance between oxygen sites in the MgO(001)
surface \modif{(5.7 $a_0$)} while the dimensions of the fivefold
ring in Na$_6$ do not fit well to the surface. That will play a role
in the dynamical evolution studied later on.  \modif{The equilibrium
distance of the lower cluster plane (facing towards the surface) and
the first surface layer is 5 $a_0$.}

\section{Results and discussion}
\label{sec:results}

\subsection{\label{chap:deposit_siteprop}Na monomer on MgO -- the
  influence of sites}

A two component system like MgO has more possible adsorption sites
than a homogeneous material like an argon substrate. The properties of
an oxygen site are much different from those of the magnesium site
because of the \modif{much larger polarizability of oxygen}.  We will
thus consider four positions with respect to the surface~: O site, Mg
site, hollow and bridge. The structure calculations of~\cite{Bae07a}
have shown that O, due to its large polarizability, is the most
attractive site while Mg acts like a repulsive site on the cluster.
This is in agreement with quantum chemical ground state calculations
of transition metals on MgO \cite{WinklerDiplom}.
\begin{figure*}[htbp]
\centerline{
\includegraphics[width=0.9\linewidth]{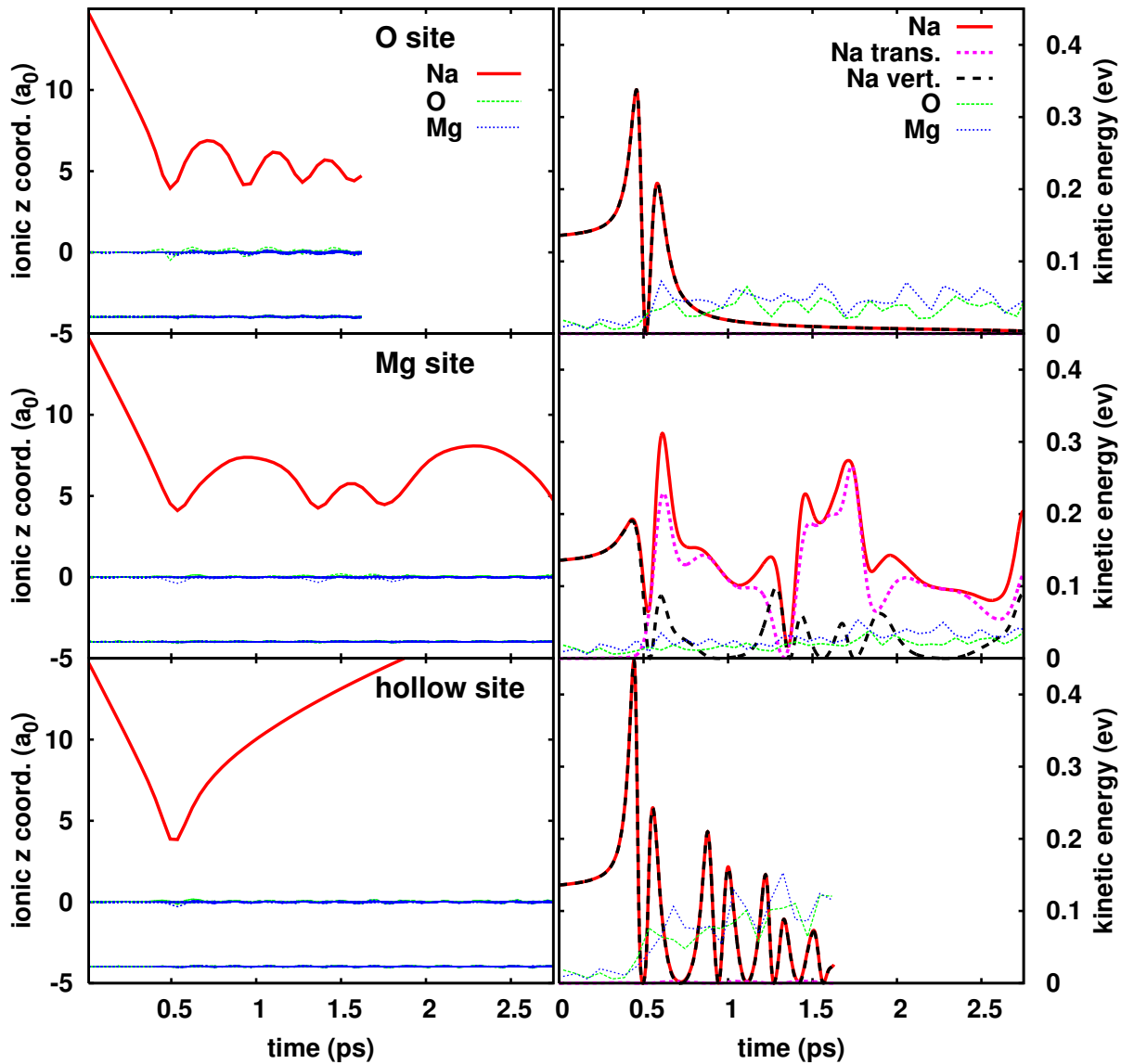}
}
\caption{\label{fig:col1M}
  Time evolution of the ionic coordinates and kinetic energies $E_{\rm kin}$
  for the deposition of a Na monomer on MgO(001) with initial
  kinetic energy $E_{\rm kin}^0=0.136$ eV, impinging on various
  sites~: O site (top), Mg site (center) and hollow site
  (bottom). Left~: $z$ coordinates of Na (thick line), Mg (thin
  curve), and O (gray line) cores. Right~: Total $E_{\rm kin}$ of Na
  (thick gray or red curve), lateral $E_{\rm kin}$ of Na (magenta
  dots), vertical $E_{\rm kin}$ of Na (black dashes), and total
  $E_{\rm kin}$ of Mg (thin dark or blue curve) and O (thin light or
  green curve) cores. 
}
\end{figure*}

In order to check the adsorption properties of the various sites, we
first study the deposition dynamics of a Na monomer. 
We briefly remind the static properties of Na@MgO(001).
The O site is most attractive, binding  Na
5 $a_0$ above the surface with energy 0.25 eV.
The Mg site is dominantly repulsive. The hollow and bridge sites
lie in between these extremes.
For deposition dynamics, the atom was initialized 15 $a_0$ away from the
substrate 
above an O site, Mg site or hollow site
respectively, each with an initial momentum along $z$ direction, 
pointing perpendicular
towards the surface 
with a magnitude corresponding to a kinetic energy
$E_{\rm kin}^0=0.136$ eV. Figure \ref{fig:col1M} shows the
results of the simulation. 
The $z$-coordinates are chosen such that the (average)
MgO surface layer resides at $z=0$.
The simplest case is the impact on the O site. The atom approaches the
surface up to a distance of $4.5\,a_0$ which is reached at about 500
fs and transfers part of its momentum to the substrate ions.  The
transfer proceeds at a very short time scale.  The surface itself is
excited mainly by the first collision which initially only affects the
ions in the immediate vicinity of the atom at closest impact. The
perturbation quickly spreads over the surface, but the associated
sound wave does not penetrate very deep into the surface. The
oscillations in the third layer are already almost
negligible. After the instant of closest contact, the atom bounces
back, but it has already lost so much energy that it cannot escape
from the surface anymore.  Thus it performs damped oscillations, with
each bounce transferring some momentum to the surface and being
practically adsorbed within the first 2\,ps.  The final distance
approaches nicely the equilibrium distance of 5 $a_0$.
The right upper column of figure
\ref{fig:col1M} shows the corresponding kinetic energy contributions. 
In the first $400\,\rm fs$, the attraction from
the MgO substrate leads to a rapid increase of the kinetic energy of
the atom up to $0.45$~eV. At the point of closest contact, the
repulsive part of the interface potential stops the atom
abruptly.
That first collision transfers by far the largest amount of 
 energy to MgO, whereas
at all subsequent collisions the energy decreases more slowly. In
order to check that the oscillations proceed only perpendicular to the
surface, the kinetic energy of the Na atom has been split into
contributions from perpendicular (or vertical) and parallel (or
transverse) motion. The latter is too small to be visible in figure
\ref{fig:col1M} and practically negligible. Thus the motion of Na proceeds
strictly perpendicular to the surface.
The kinetic energy transferred to the MgO can also be read off from
figure \ref{fig:col1M} (see right upper panel).  The contributions
from oxygen and magnesium are given separately.  Oxygen ions are the
lighter species and therefore react first being quickly accelerated.
About 100 fs later, the energy has already been distributed almost
equally over both ion types.

The dynamics behaves totally different if the atom impinges on the 
(repulsive) Mg site, see middle panels of figure \ref{fig:col1M}.  At
first glance, 
the $z$-component of the Na trajectory looks quite similar to the case
before.  But one notes that the motion is not damped after the first
reflection.  The kinetic energies (middle right panel) give a clue on
the process. 
There is much less energy transfer at first impact which is related to
the fact that the Mg$^{2+}$ ion is more inert. And there is a
significant amount of lateral kinetic energy for the Na atom creeping
up after impact time at 500 fs.  In fact, most of the kinetic energy
is now in lateral motion.  The atom is deflected by the Mg$^{2+}$
ion. \modif{ It is to be noted that the annealing of the substrate
configuration leaves a small amount of symmetry breaking with
fluctuations of the atomic positions of about 0.05 a$_0$.  This small
symmetry breaking allows the atom to acquire sidewards momentum and so
it} bounces away in sideward direction, hops over the surface several
times changing direction whenever it comes close to another surface
ion. The motion is almost undamped because little energy is
transferred to the surface after the first collision. The atom has
thus still too much energy to be caught by a certain site of the
surface.  But as the atom cannot escape the surface as a whole, it
will continue to lose slowly energy and finally be attached to an
oxygen site, long after the simulation time of 3 ps.

The bottom panels of figure \ref{fig:col1M} show the case of impact at
an hollow site. We see again the immediate reflection at impact time
associated with fast energy transfer. Less energy is transferred 
than on the other sites (see upper and middle panels)
and thus the bounce-back has a much larger amplitude than in both
other cases. The Na motion remains strictly perpendicular to the
surface as practically no
lateral kinetic energy can be seen. The vertical kinetic energy is
almost approaching zero because the departing Na atom has to work
against the polarization potential. The case is at the limits of our
box size and energy resolution such that we cannot decide whether the
atom will finally escape with extremely small kinetic energy, or
will bounce back and relax to an adsorption site on a very
long time scale. Nevertheless, we find it worthnoting that the hollow
site seems sufficiently attractive to hinder deflection towards the
still more attractive oxygen site.

\subsection{\label{chap:deposit_basictrends8}
Cluster deposition }

\subsubsection{The case of symmetric Na$_8$}
\begin{figure}[htbp]
\centerline{
\includegraphics[width=\linewidth]{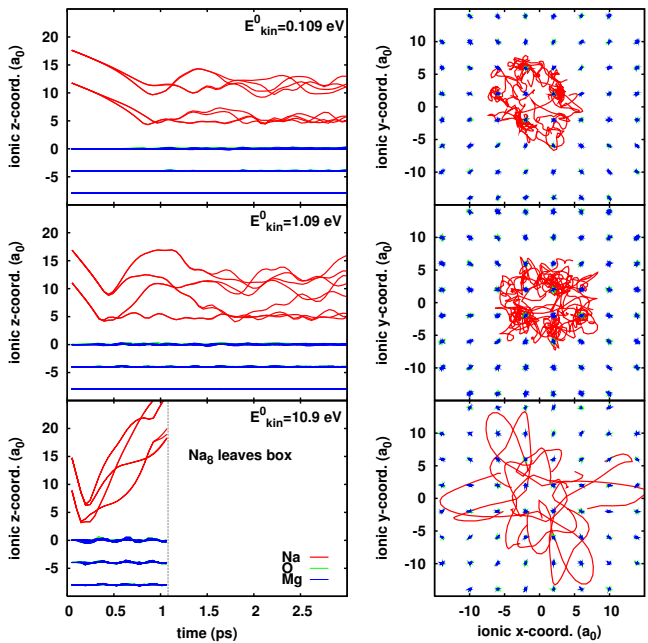}
}
\caption{\label{fig:traj8}
Left~: Time evolution of the ionic $z$-coordinates of Na$_8$ approaching
MgO(001). Right~: Projection of the Na and MgO trajectories into the
$x$-$y$ plane; the figures are vertically sorted by increasing initial
kinetic energy~: Soft deposition (top), robust deposition (center),
reflection (bottom).
}
\end{figure}
The analysis of section \ref{chap:deposit_siteprop} has shown the
importance of surface site nature in the deposition
process. Depositing an extended object such as a cluster will lead to
a mixed situation because the ions of the cluster will necessarily be
placed above different sites. In the following, we will discuss
deposition of Na$_8$ and Na$_6$ which have very different structures
and so promise to show different deposition scenarios.
As a first step in the analysis, we shall consider 
detailed ionic trajectories both perpendicular 
and parallel to the surface.
The case of Na$_8$ is shown in figure \ref{fig:traj8}. 
The cluster was injected with its symmetry axis pointing
through a hollow site and with the lower ring facing closer to
bridge sites.
The top panels show a soft deposition where
the initial kinetic energy is $0.109$~eV ($0.0136$~eV per
Na ion).  The left upper panel shows that the cluster is slightly
accelerated in the initial phase, due to the attraction from the
surface. But that acceleration differs for the different ions on the
lower ring because they approach different sites on the surface.  At
the same time, the cluster rotates in the $x$-$y$ plane to bring the
four ions of the lower ring closer to the attractive oxygen sites. One
may spot that from the top view in the right upper panel.
At the point of closest impact around 900 fs, the cluster
transfers some momentum to the surface. The substrate ions are
slightly displaced from their equilibrium positions and oscillate
around their new positions. The disturbance quickly decreases from
layer to layer. The perturbation is negligible already in the fourth layer.
%
The complicated detailed dynamics of the Na ions indicates that a
major part of the translational kinetic energy is converted into heat,
i.e.\,kinetic energy of the motion relative to the center of mass, as
will be confirmed in section \ref{chap:deposit_energytransfer}.
Nevertheless, the cluster basically keeps its original structure
during the whole simulation period of 9 ps. In particular the two
rings, each made of four ions, always stay clearly separated from each
other. The top-down projection of the trajectories (see right-hand
side of figure \ref{fig:traj8}) shows that the cluster as a whole (or
its center of mass) remains oscillating around the point of impact. The
remaining kinetic energy of the cluster does apparently not suffice to
overcome the surface corrugation barriers.  This is related to the
fact that the Na$_8$ structure fits approximately well to the
structure and binding distance of MgO, see section \ref{sec:struct}.

The middle panels of figure~\ref{fig:traj8} show a more robust
deposition dynamics with initial kinetic energy
$E_\mathrm{kin}^0=1.09$~eV.  The initial velocity is higher and
the impact time comes earlier, now at 450 fs.  The pattern remains, in
principle, similar to the softer deposition.  There is little momentum
transfer to the substrate, strong internal excitation of the cluster,
and the cluster is not departing too far from the impact
point. However, perturbations are much larger, yielding larger
amplitudes in vertical and lateral motion.  As a consequence, the
two rings of Na$_8$ are now not always clearly separated.
Nevertheless the typical structure of Na$_8$ reappears from time to
time as we will see later.
The top-down projection of the trajectories (middle right panel of
figure \ref{fig:traj8}) indicates a new effect,
a sideward drift from one adsorption site to the next equivalent
site.  This sideward drift is again induced by the interplay between
attractive O and repulsive Mg sites. The chaotically moving Na ions
explore a strongly corrugated surface which leads to occasional side
kicks from the repulsive Mg sites.

The bottom panels of figure \ref{fig:traj8} show a hard collision with
initial kinetic energy $E_\mathrm{kin}^0=10.9$~eV.  Internal
cluster and excitation and surface perturbation are, of course, again
larger. The new feature is that the cluster is reflected from the
surface and leaves the numerical box at about 1 ps, however with huge
internal excitation. It is not clear whether the departing cluster
will stay asymptotically stable. That is beyond our simulation
capacity.

\begin{figure}[htbp]
\centerline{
\fbox{
\includegraphics[width=8.1cm]{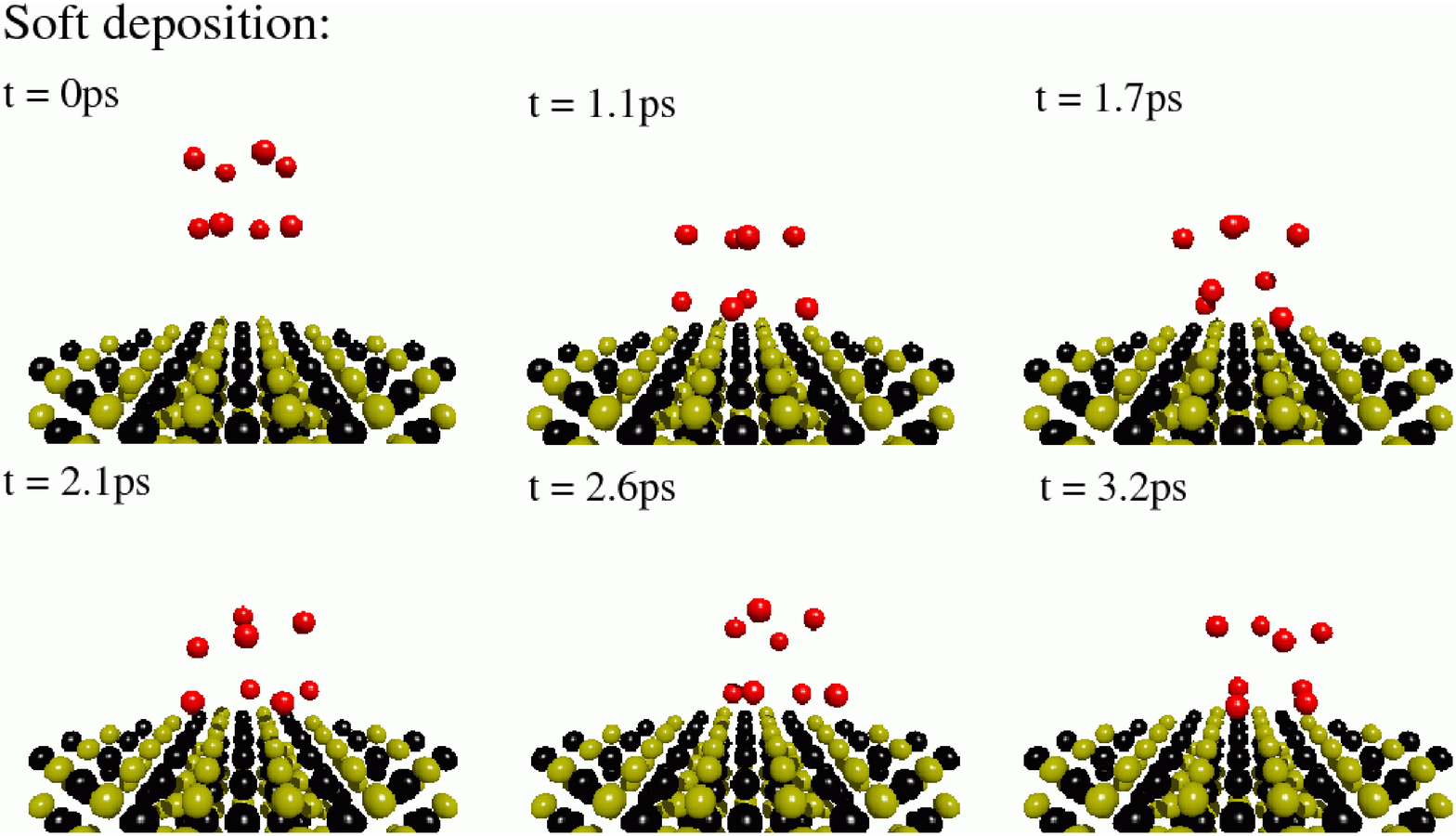}}
}
\medskip
\centerline{
\fbox{
\includegraphics[width=8.1cm]{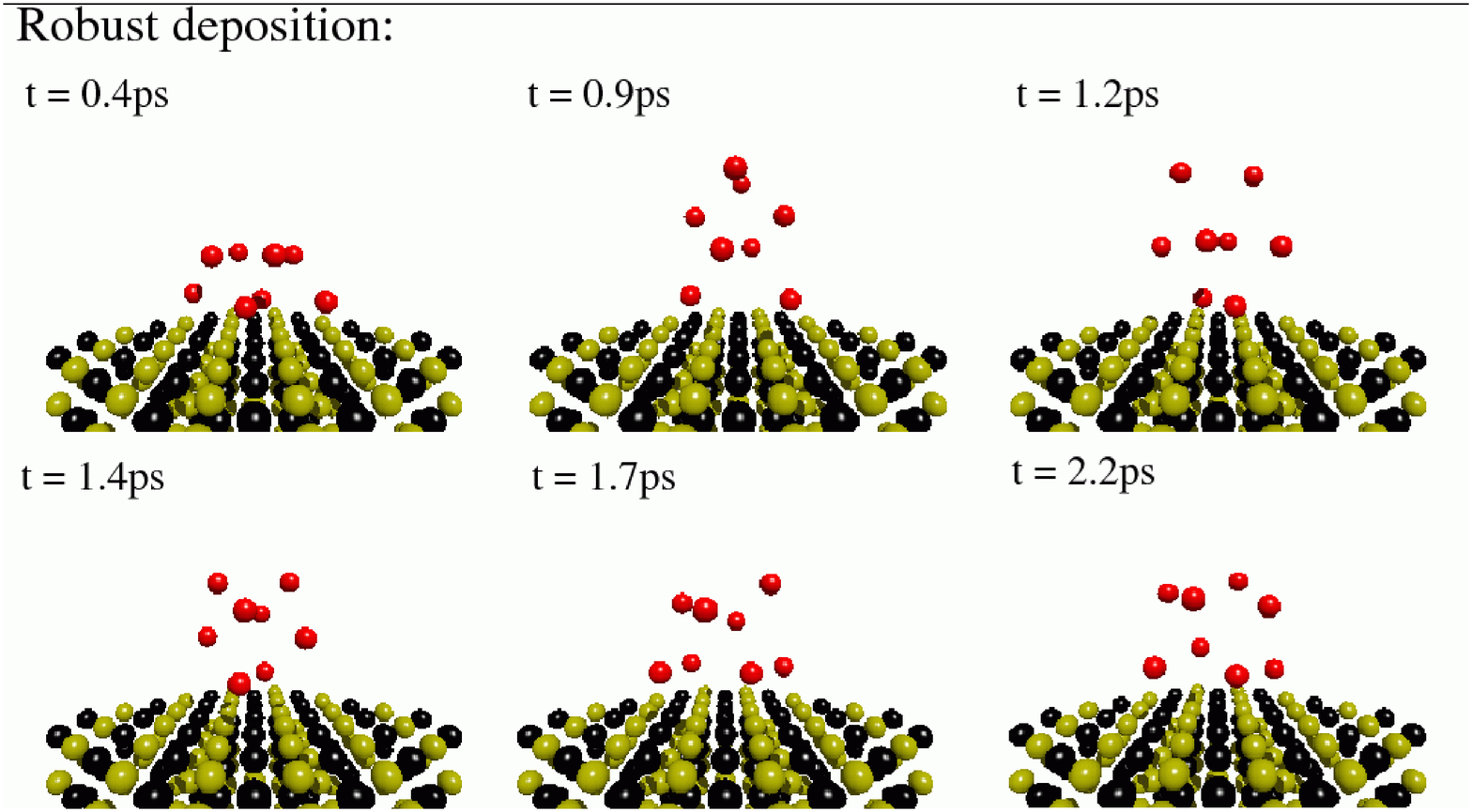}}
}
\medskip
\centerline{
\fbox{
\includegraphics[width=8.1cm]{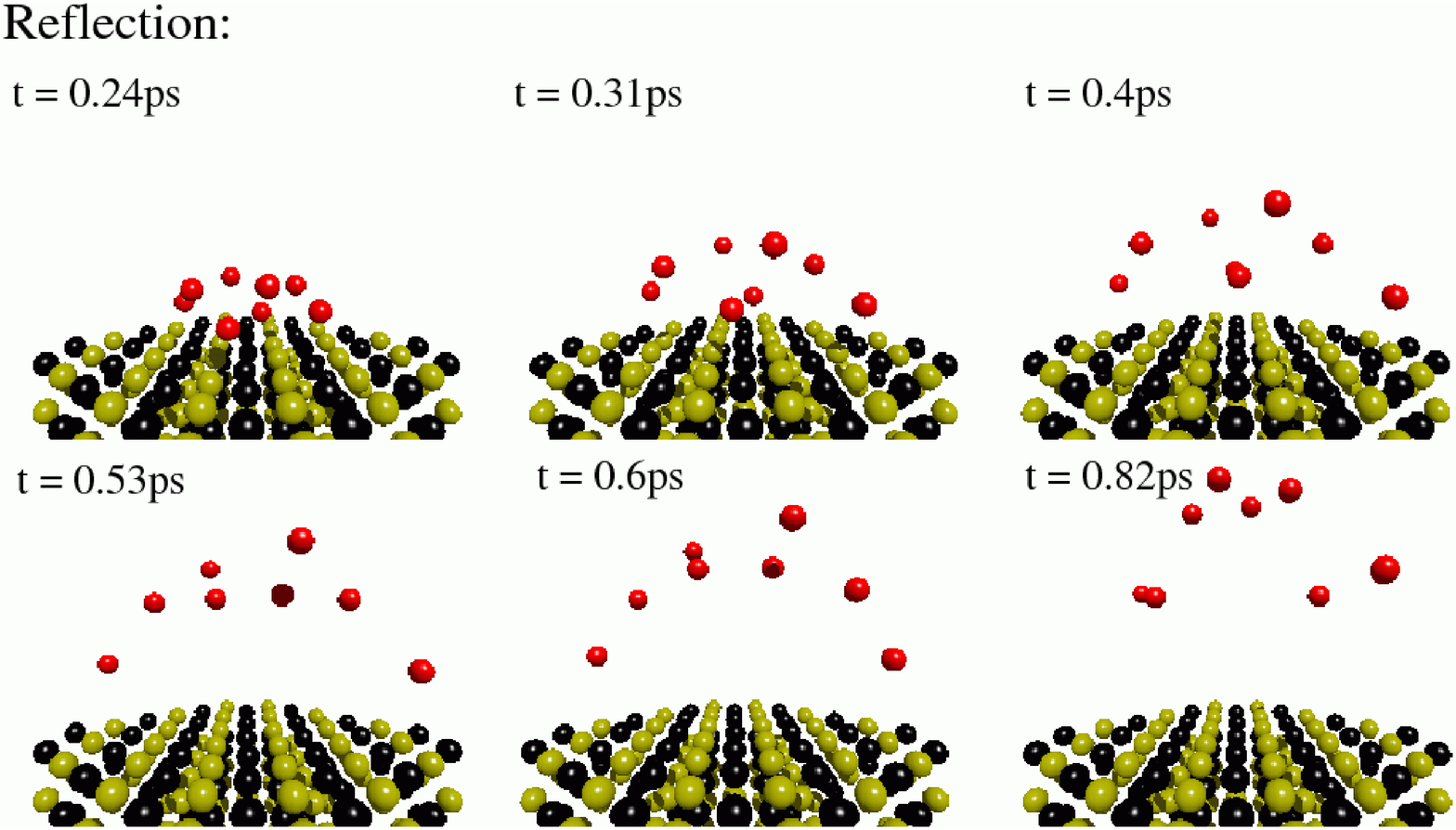}}
}
\caption{\label{fig:snapshot}
Snapshots of the time evolution 
for three cases of collisions of Na$_8$ on MgO(001)
with different initial kinetic energies $E_{\rm kin}$.
Upper panel~: Soft deposition with $E_{\rm kin}^0=0.109$~eV.
Middle panel~: Robust deposition with $E_{\rm kin}^0=1.09$~eV.
Lower panel~: Reflection with $E_{\rm kin}^0=10.9$~eV.}
\end{figure}
Figure \ref{fig:snapshot} complements the view within showing a
sequence of snapshots of the detailed structure for each of the
three cases discussed above. The uppermost panel for soft deposition
nicely shows the initial rotation of the cluster to match the
attractive oxygen sites. The further snapshots indicate the sizeable
internal excitation, however remaining small enough to see at all
times clearly the two-ring structure of Na$_8$.
The middle panel for more robust deposition also presents the
much larger cluster oscillations where the original cluster structure
is often blurred, but reappears shortly at other times. That
demonstrates the surprisingly good binding of Na clusters,
particularly the Na$_8$ cluster with its magic electron configuration.
The lowest panel shows the case of reflection. Obviously, some ions
would like to stick to the surface, but are finally caught back by the
cluster which departs in a highly excited state.

\subsubsection{The case of strongly oblate Na$_6$}

\begin{figure}[htbp]
  \centerline{
\includegraphics[width=\linewidth]{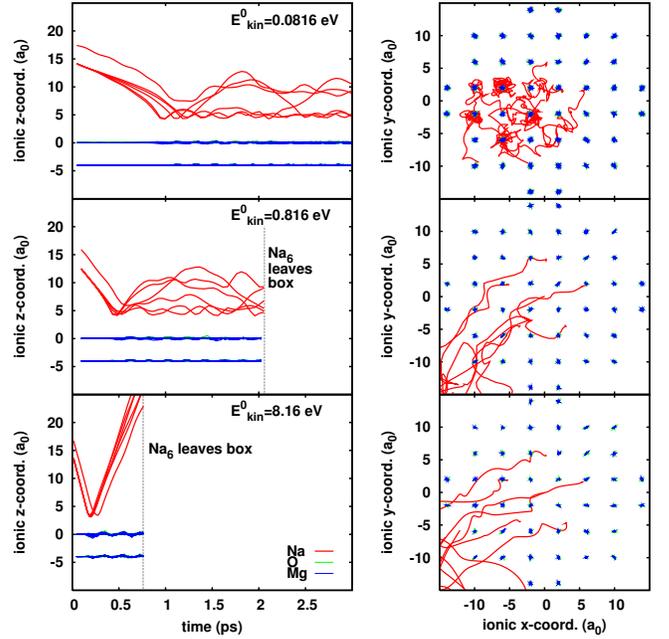}
}
\caption{\label{fig:traj6}
Left~: Time evolution of the ionic z-coordinates of Na$_6$ approaching
MgO(001). Right~: Projection of the Na and MgO trajectories into the
x-y-plane; the figures are vertically sorted by increasing initial
kinetic energy~: Soft deposition (top), robust deposition (center),
reflection (top).
}
\end{figure}
Results for the deposition of Na$_6$ are shown in figure
\ref{fig:traj6}. The impact energy is varied and all three cases start
from the same initial configuration where the top ion of Na$_6$ (see
section \ref{sec:struct}) is facing away from the substrate and the
fivefold ring is parallel to the surface.
The general features are similar to the case of Na$_8$. One observes  a
large internal excitation of the cluster while comparatively little
perturbation goes to the substrate and there is again the clear
distinction between deposition for lower impact energies and
reflection for higher ones.  But there are several interesting
differences in detail. Most of all, there is a strong lateral drift
in all cases. Indeed the pentagonal ring of Na$_6$ does not match the
rectangular structure of MgO, which hinders it from fully accomodating
the attractive oxygen sites. Thus one or two corners of the pentagon
are bent up during the deposition process, and this, in turn, induces a
sizeable lateral momentum (see right panels of figure
\ref{fig:traj6}), and a strong perturbation of the pentagon, as can be
deduced from the motion of $z$-coordinates shown in the left panels.
In the case of the robust deposition, the cluster even rolls over the
surface. The stronger lateral excitation 
leaves also somewhat more perturbation to the substrate than in the
case of Na$_8$, as may be spotted when comparing figures
\ref{fig:traj6} and \ref{fig:traj8}. That will become more obvious
when checking energies in section \ref{chap:deposit_energytransfer}.
Finally, it is interesting to note that the cluster orientation is also
reverted in the case of reflection.  The $z$ coordinates (upper left
panel in figure \ref{fig:traj6}) suggest a process where the ring and the
former top ion are reflected "independently" such that the topping ion
is departing "behind/after" the ring, and thus reverting the cluster
orientation.

\begin{figure}[htbp]
\centerline{
\includegraphics[width=0.85\linewidth]{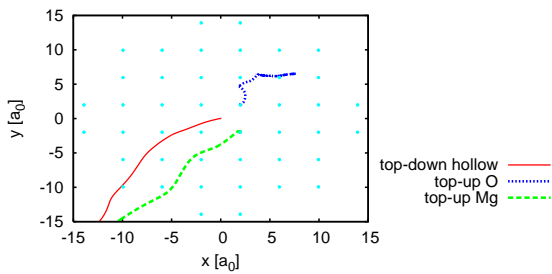}
}
\caption{\label{fig:lateraldrifts}
The trajectories of the Na$_6$ center of mass, projected on a plane
parallel to the surface. The lateral drift sensitively depends on the
site above which the cluster impinges (hollow, O or Mg site) and on
its orientation (top ion above -- up -- and top ion below -- down --
the ring).
}
\end{figure}
As noted above, when deposited, the Na$_6$ experiences a sizable 
drift due to the mismatch of its structure with the crystalline structure 
of MgO. One thus expects that direction and strength of the lateral motion depend
sensitively on the initial position and orientation of Na$_6$ relative
to the surface. Indeed figure \ref{fig:lateraldrifts} shows the
trajectory of the center-of-mass of the Na$_6$ cluster projected onto
the $x$-$y$-plane for three different initial orientations. There 
are obvisouly dramatic differences. The cluster is kicked to a
strong lateral motion for initial impact at repulsive sites (Mg,
hollow) while only moderate lateral drift appears for impact on the
attractive O site.

\modif{One can learn more about the electronic charge distribution in
the cluster during a collision by a direct multipole analysis of this
distribution. We discuss here briefly the lowest non-trivial
moments, the dipoles.
\begin{figure}[htbp]
\centerline{
\includegraphics[width=0.98\linewidth]{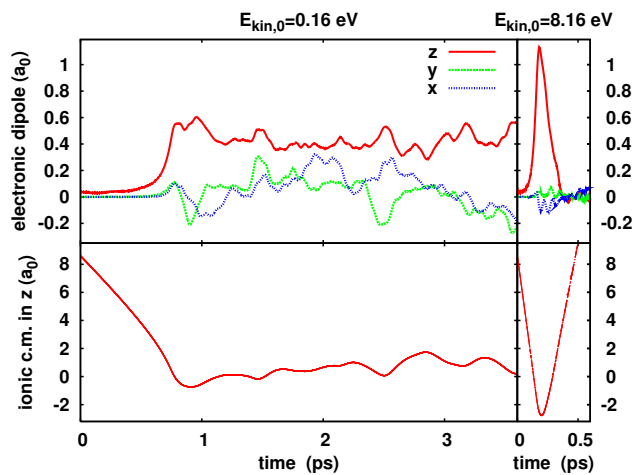}
}
\caption{\label{fig:Na6MgO-dip-CM}
Dipole polarization of the cluster electrons during collision of
Na$_6$ on MgO(001). 
Upper panels: Dipole moments of the electron cloud of
the Na$_6$ cluster in $x$-$y$- (horizontal) and
$z$-direction as function of time.
Lower panels: Time-evolution of the $z$ componente of the
center-of-mass of the Na$_6$ cluster; the line $z=0$ corresponds to the center of mass
of a fully relaxed configuration of Na$_6$ on MgO(001).
Left panels: case of soft deposition at rather low initial
kinetic energy (as indicated).
Right panels: case of more energetic collision which leads to
immediate reflection. 
}
\end{figure}
Figure \ref{fig:Na6MgO-dip-CM} shows the time evolution of dipole
polarization for two different scenarios, deposition
vs. reflection. The figure is augmented by the time-evolution of the
center-of-mass as global indicator of the dynamical situation (lower
panels).  The two scenarios differ by the initial kinetic energies,
the lower value related to a more or less soft deposition, while the
higher initial velocity leads to immediate reflection of the cluster,
as discussed above.  In the slow deposition process (left panels),
there is a strong increase of the $z$-polarization at the time of
closest impact. This polarization remains during the further evolution
at about 10\% of the Wigner-Seitz radius, hence represents a
considerable internal polarization. Also, some $x$-$y$ polarization
builds up during the ongoing deposition oscillations. In contrast, in
the case of a reflection (right panels), one sees a large
instantaneous polarization at the time of closest approach, but only a
very small remaining effect when the cluster has departed from the
surface. The very short interaction time limits the internal
excitation.  }

\subsection{MgO versus Ar Substrate}
\label{chap:deposit_MGOvsAR}

In previous works~\cite{Feh06a,Din07a}, the deposition of Na$_6$ on a
cold, condensed argon substrate Ar(001) was investigated.  Like MgO,
solid Ar has a large band gap.  But apart from this insulating nature,
the two materials have much different properties.  The attractive
interaction between MgO and Na is much stronger than between Ar and
Na, due to the larger polarizability of the oxygen ion.  On the other
hand, frozen Ar material is a very soft solid due to the weak Ar-Ar
binding. The melting point of Ar is $83.78\,\rm
K$~\cite{MeltingArgon}, much lower than that of MgO around $3073\,\rm
K$~\cite{Melting}. The softness of the
Ar material thus changes the energy 
balance to the extent that the Ar substrate takes up most of the
impinging energy, leaving rather little internal excitation for the
cluster itself.  Thus Ar substrates are very efficient soft stopper materials.
In the former analysis \cite{Feh06a,Din07a}, we had run a similar
series of impact energies as above and we also found soft deposition
for energies up to at least $E_{\rm kin}^0/N_\mathrm{ion}=0.272$~eV.
An attempt to reach a  reflection regime by further increasing the
impact energy then led to a significant destruction of the substrate.
\begin{figure}[hbtp]
\includegraphics[width=0.8\linewidth]{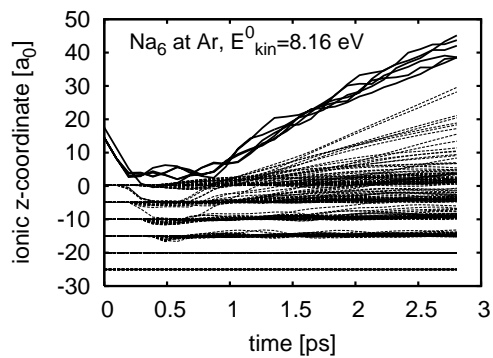}
\caption{\label{fig:depositArgon}
Time evolution of $z$ coordinates during collision of
Na$_6$ with initial total kinetic energy
$E_{\rm kin}^0=8.16$~eV (1.36 eV per Na atom) on an Ar(001) surface.
 Data taken from \cite{Feh06a}.
}
\end{figure}
Figure \ref{fig:depositArgon} illustrates that violent collision of
Na$_6$ on an Ar(001) surface. The cluster is indeed finally reflected,
but the process evolves much different from the case of the collision
with MgO shown in figure \ref{fig:traj6}.  The cluster is not
reflected instantaneously, as for MgO, but with a delay of about 500
fs.  It requires an additional boost from momentum reflected by the
first Ar layer to finally release the cluster. Moreover, so much
energy has been deposited in the weakly bound Ar material that the
substrate is seriously damaged by that forced ``reflection''.  These
significant differences between Ar and MgO substrate will also be seen
in the energy analysis later on.

\subsection{\label{chap:deposit_energytransfer}Energy Transfer}

\subsubsection{Time evolution of energy components}
\begin{figure}[htbp]
\includegraphics[width=0.8\linewidth]{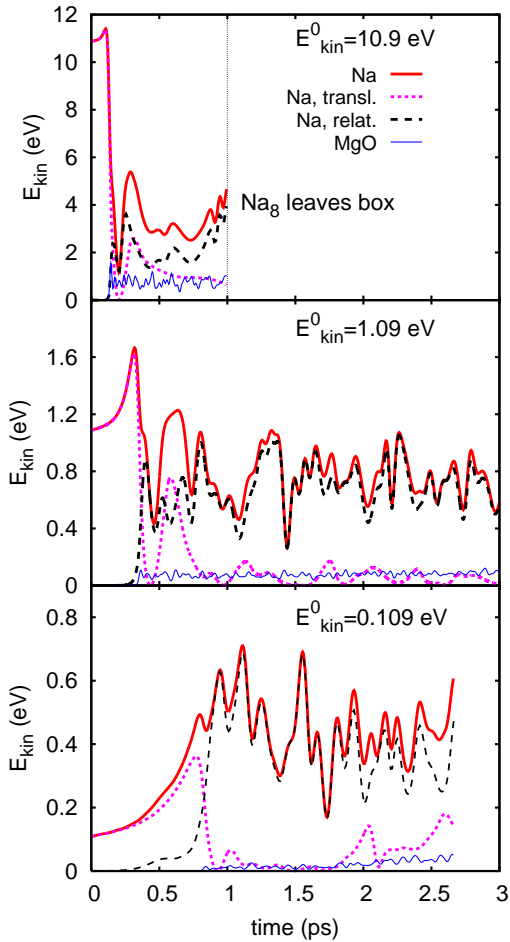}
\caption{\label{fig:kinen8}
Time evolution of various contributions to kinetic energy 
for the collision of Na$_8$ with MgO(001) at three different
initial kinetic energies as indicated. Contributions are~:
Total kinetic energy of the cluster ions (Na, thick curves),
contributions due to center of mass motion (Na transl., dashes)
and relative motion or heat (Na relat., dots), and total kinetic
energy of the substrate (MgO, thin curves). 
}
\end{figure}
A complementing view of the deposition dynamics is given by the
kinetic energies. Figure \ref{fig:kinen8} shows the time evolution of
the kinetic energies for Na and MgO. The kinetic energy for the Na
cluster is furthermore splitted into center-of-mass energy and
intrinsic kinetic energy (from the motion relative to the
center-of-mass).
Let us first consider the case of reflection (upper panel).  In the
approaching phase, the cluster is accelerated by about 0.68~eV which
is small compared to the initial energy 10.9~eV.  Dramatic and fast
changes emerge at impact time at 200 fs. The cluster kinetic
energy exhibits a deep minimum. In that stage, almost all energy is stored in
deformation. A large part of that deformation energy is quickly released
showing up now as intrinsic kinetic energy of the cluster plus a
smaller bit in translational energy. Another small fraction of energy is
transferred to the substrate. The translational kinetic energy
decreases further on, because the departing cluster has to work against
the attractive polarization interaction. Still, there remains
sufficient translational energy to allow the cluster to finally
escape,  as in a very inelastic collision.

Similar results are found for the soft (lower panel in figure
\ref{fig:kinen8}) and robust deposition (middle panel).  Again, only a
small fraction of the energy is transferred to the substrate, another
small fraction goes to the cluster center-of-mass oscillations, and
the major part to intrinsic energy of the cluster.
Particularly interesting is the case of robust deposition (middle
panel) where it requires a second bounce to stir up intrinsic cluster
motion. Before that, there is still enough energy in translation to
allow a lateral hopping from one attractive MgO site to the next
(see also figure \ref{fig:traj8}).
It is worthnoting that the average trend of the kinetic energy
of MgO in figure \ref{fig:kinen8} has a small, but nonvanishing,
slope. The cluster continues to exchange energy with the substrate on
a very slow pace. That indicates a thermalization process which
eventually leads to equidistribution of kinetic energies after long
time, however much beyond our simulation capabilities.

\begin{figure}[htbp]
\includegraphics[width=0.8\linewidth]{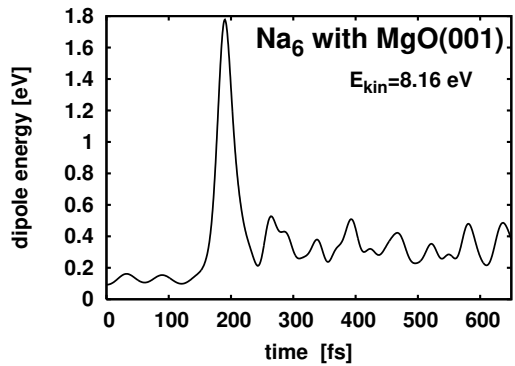}
\caption{\label{fig:Na6-MgO-dipole-energies}\modif{
Time evolution of the dipole polarization energy in the 
MgO(001) substrate for the collision for Na$_6$ with MgO(001)
at an initial kinetic energy of 8.16 eV (reflection case).}
}
\end{figure}
\modif{The present modeling includes an independent dynamics of the
dipole moments of the oxygen anions in the substrate. 
It has been shown recently that a significant amount of energy
can be stored in these degrees-of-freedom when a metal cluster is
deposited on an Ar surface\cite{Din08a,Din09}. One can define a dipole
energy which scales as the square of the dipole amplitudes.}
Figure
\ref{fig:Na6-MgO-dipole-energies} shows a typical result for the time
evolution of the energy contained in the oscillating dipoles
\modif{in the case of reflection of Na$_6$ deposited on MgO}. 
There
is a small initial value which corresponds well to the finite initial
distance of the Na cluster to the substrate. There is a large
contribution at the time of closest impact. This part is dominated by
(instantaneous) static polarization which would also be contained in a
Born-Oppenheimer MD. The dipole energy falls back to lower values when
the cluster departs from the substrate (see figure
\ref{fig:traj6}). But there remains some offset which corresponds to
the energy finally transferred to the dipole degrees of freedom. It
amounts to about 2\% of the impact energy, \modif{which is small
compared to the other energetic observables, see
figure~\ref{fig:phases68} and corresponding discussion in
section~\ref{sec:redistrib}}. \modif{In contrast to the case of Ar
substrate, energy transfer to MgO dipoles} has
actually only a 
small effect for the overall ionic dynamics. But more subtle
properties as optical response and trajectories of free charges
(to be discussed in a subsequent publication) will be
sensitive to such details.

\subsubsection{Energy transfers "at" impact}

Notwithstanding asymptotic thermalization, the fast energy transfer to
the substrate in the early stages is an interesting observable
characterizing the collision process.  %
\begin{figure}[htbp]
\includegraphics[width=0.7\linewidth]{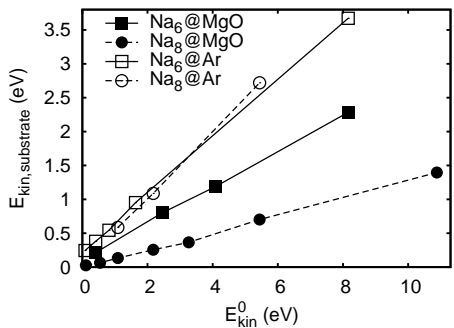}
\caption{\label{fig:surftemp}
Heating of the MgO or Ar substrate after impact of Na$_6$ and Na$_8$.
The energy transfer to Ar is independent of the cluster structure,
whereas the energy transfer to MgO is not. Na$_6$ transfers about
twice as much energy as the compact Na$_8$.
}
\end{figure}
Figure \ref{fig:surftemp} shows the kinetic energy of the substrate
soon after the collision, i.e. averaged over the first 2 ps after
impact, as a function of the initial kinetic energy of the cluster
$E_{\rm kin}^0$.  Apparently the energy absorbed by the substrate is
proportional to $E_{\rm kin}^0$. But the slope depends very much on
cluster and surface types.  The soft Ar substrate absorbs much more
energy than MgO, typically a bit more than 50\% of the initial kinetic
energy.  The softness and the rather small surface corrugation of Ar
make the process insensitive to the actual cluster which is
approaching. That is different for MgO. There is always less energy
absorption by the substrate and there is a strong dependence on the
cluster configuration.  Na$_6$ transfers more than twice as
much energy as Na$_8$. The strong surface corrugation of MgO induces
that sensitivity to cluster geometry. Remind that Na$_6$ does not
match very well to the MgO surface while Na$_8$ does (see section
\ref{sec:struct}).

\begin{figure}[htbp]
\includegraphics[width=0.7\linewidth]{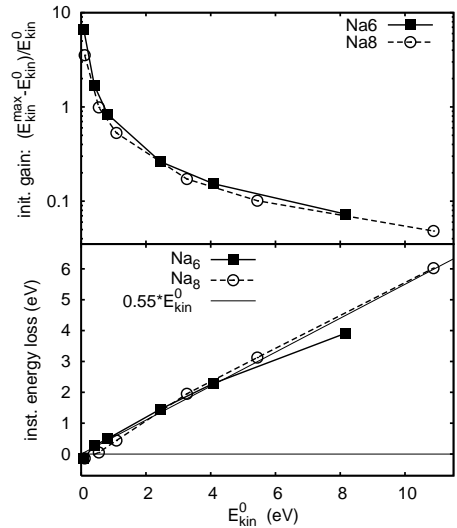}
\caption{\label{fig:Egain}
Upper panel~: The gain in kinetic energy of the cluster when
approaching the MgO surface; lower panel~: Instantaneous energy loss,
defined as the difference between maximum kinetic energy before the
impact and next maximum after impact.
}
\end{figure}
More information about what is happening directly around impact time
can be obtained by reading off observables at shorter time scales
(shorter than the 2 ps used above).  Figure \ref{fig:Egain} shows two
such observables as a function of initial kinetic energy.  The upper
panel shows the energy gain in the approaching phase due to the
acceleration by the polarization potential. It is defined as the
difference of the first maximum of the cluster kinetic energy and the
initial energy. 
Na$_6$ acquires slighlty more energy than Na$_8$ because it has a
non-vanishing dipole moment which, in turn, enhances the polarization
attraction.  There is a large gain in the low energy range (the regime
of soft deposit), while the trend becomes very flat for fast collisions.  
This is probably due to a move from adiabatic to non-adiabatic relaxation
processes in the surface. For very low impact velocities, the surface
ions have time to follow the forces from the cluster, whereas for very
high velocities the surfaces ions do not have enough time to respond
before the cluster collides.
The lower panel of figure \ref{fig:Egain} shows an attempt to quantify
an ``instantaneous energy loss''. To that end, we take difference
between the maximum kinetic energy before the impact and the next
maximum after the impact. Obviously the instantaneous energy transfer
is practically the same for Na$_6$ and Na$_8$ independent of their
differences in structure; and the energy loss is approximately
proportional to $E_{\rm kin}^0$. About half (more precisely around 55\%)
of the impact energy is withdrawn from the cluster in that first round.

\subsubsection{Redistribution of initial kinetic energy}
\label{sec:redistrib}

\begin{figure}[htbp]
\centerline{
\includegraphics[width=0.7\linewidth]{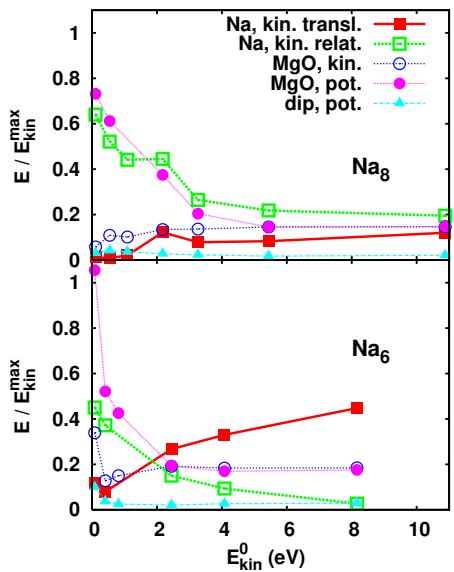}
}
\caption{\label{fig:phases68} 
Distribution of impact energy amongst various components~: 
Kinetic energy of the cluster (split into intrinsic, open squares, and
translational, close squares, motion), kinetic energy of the MgO
substrate (open circles), potential energy in the 
oxygen dipoles (triangles), and other potential energy in the
substrate (close circles).
Upper panel~:
Na$_8$ with MgO, lower panel~: Na$_6$ with MgO.  }
\end{figure}
It is, furthermore, interesting to see how the initially available
energy is distributed over the various constituents.  Such energy
balance is shown in figure \ref{fig:phases68} as a function of $E_{\rm
kin}^0$.  All energies have been averaged over 2 ps after impact
time. They are drawn relative to the maximum kinetic energy of the
cluster before impact.  The energy terms do not
necessarily sum up to one because the interaction between substrate
and cluster is omitted, as well as the intrinsic potential energy of
the cluster. Low energies ($< 2.72$~eV for Na$_8$ and
$<2.18$~eV for Na$_6$) represent the regime of deposition. The largest
amount of energy is used up here for intrinsic cluster motion combined
with potential energy of the substrate. The latter is responsible for
the attachment of the cluster to the surface. The higher energies 
represent the regime of reflection. The share of energies depends 
here on the cluster geometry.  For Na$_6$, the translational motion
takes the lead, whereas the intrinsic motion shrinks to almost zero.
That complies with the trajectories in figure \ref{fig:traj6} which
show that the cluster is repelled from the surface with opposite
orientation but only weakly perturbed structure.
The reflection of the top ion seems to proceed independently from the
five-fold ring. The ring hits first and departs first while the top
ion is reflected later thus departing behind the ring.

For both clusters, kinetic and potential energies of the MgO behave
similar.  For high $E_{\rm kin}^0$ in the reflection regime, potential
and kinetic energy are equal. The cluster quickly transfers 
some momentum to the substrate at impact and then disappears.  This
leaves the substrate ions oscillating around their equilibrium
positions in harmonic motion, associated to equipartition 
between kinetic and potential energy.  For high $E_{\rm kin}^0$ in the
soft landing domain, the situation is different.  The potential energy
becomes the dominant contribution because the cluster is adsorbed and
remains in contact with the surface. This distorts the surface and
leads to a large potential energy.
Polarization energy is dominating in the potential energy.  But the
isolated contribution from the oxygen dipoles, also shown in figure
\ref{fig:phases68}, is comparatively small.  The
polarization within the oxygen ions in fact remains  small as compared to the
polarization caused by the displacement of O versus Mg, each one
carrying a net charge of $\pm 2e$.

\section{Conclusions}
\label{sec:concl}

We have analyzed the dynamics of deposition of small Na clusters on
an MgO surface taking up a well tested hierachical QM/MM modeling where
the cluster electrons are treated quantum mechanically by
time-dependent local-density approximation and the cluster ions as
well as the substrate atoms by classical molecular mechanics. The
dynamical polarizability of the substrate atoms is taken into account
to describe correctly the strong polarization effects in the
cluster--material interaction. The results are compared to deposition
on Ar surface which is much softer than MgO.

Test cases were Na$_6$, which is a strongly oblate cluster with a
finite dipole momentum, and the well bound and highly
symmetrical Na$_8$.  The general pattern are similar~: The clusters are
very quickly stopped by the substrate, they transfer a rather small
amount of energy to the substrate while acquiring strong internal
excitation. For larger impact energies, the clusters are reflected
from the surface.  This reflection is, of course, inelastic and leaves
the clusters departing in highly excited intrinsic motion.  There
are, on the other hand, significant differences between the two
cases, since the cluster geometry has a large influence on the
dynamics. The main effects come from the strong surface corrugation of
MgO(001). Na$_6$ does not match  the surface structure and thus
acquires significant lateral motion in contrast to Na$_8$ which keeps
better on a vertical track.  Moreover, Na$_6$ transfers more energy to
the substrate than Na$_8$.

In comparison to Ar(001) surface, we find a similar energy range for
deposition and reflection. The details, however, differ
dramatically. Deposition on Ar(001) transfers most of the energy to
the substrate leaving a rather mildly excited cluster on the surface
while there is very little energy transfer to MgO(001) and large
intrinsic excitation of the cluster. Reflection from Ar(001) is
achieved at the price of severe surface destruction while MgO remains
intact at the danger that the highly excited departing cluster may
fragment later on.

The detailed energy balance differs in the deposition and reflection
regime. In case of deposit, most energy is going to intrinsic cluster
excitation and substrate polarization. In case of reflection, 
there is, of course, more translational energy left for the cluster
and the substrate develops equipartition of kinetic and potential
energy related to the remaining small, nearly harmonic, oscillations.

\section*{Acknowledgements}
This work was supported by the {Deutsche Forschungsgemeinschaft (RE
322/10-1, RO 293/27-2),} Fonds der Chemischen Industrie (Germany), a
Bessel-Humboldt prize, and a Gay-Lussac prize.

\bibliographystyle{apsrev}
\bibliography{depos}

\end{document}